# DOES COURIER GENDER MATTER? EXPLORING MODE CHOICE BEHAVIOUR FOR E-GROCERIES CROWD-SHIPPING IN DEVELOPING ECONOMIES

**Oleksandr Rossolov[1], Anastasiia Botsman[2], Serhii Lyfenko[2], Yusak O. Susilo[1]**

[1]Institut für Verkehrswesen (IVe), University of Natural Resources and Life Sciences,
Peter-Jordan-Straße, 82, 1190 Wien, Austria
[2] Department of Transport Systems and Logistics, O. M. Beketov National University of Urban Economy in Kharkiv, Marshal Bazhanov Street, 17, 61002 Kharkiv, Ukraine

email: oleksandr.rossolov@boku.ac.at[*]; botsman.nastya@gmail.com; lyfenkosergii@gmail.com;
yusak.susilo@boku.ac.at;

O. Rossolov ORCID: 0000-0003-1495-0173, A. Botsman ORCID: 0000-0002-0308-0339,
S. Lyfenko ORCID: 0000-0002-9418-3033, Y.O. Susilo: 0000-0001-7124-7164,

**Abstract**
This paper examines the mode choice behaviour of people who may act as occasional couriers to provide crowd-shipping (CS) deliveries. Given its recent increase in popularity, online grocery services have become the main market for crowd-shipping deliveries' provider. The study included a behavioural survey, PTV Visum simulations and discrete choice behaviour modelling based on random utility maximization theory. Mode choice behaviour was examined by considering the gender heterogeneity of the occasional couriers in a multimodal urban transport network. The behavioural dataset was collected in the city of Kharkiv, Ukraine, at the beginning of 2021. The results indicated that women were willing to provide CS service with 8% less remuneration than men. Women were also more likely to make 10% longer detours by car and metro than men, while male couriers were willing to implement 25% longer detours when travelling by bike or walking. Considering the integration of CS detours into the couriers' routine trip chains, women couriers were more likely to attached the CS trip to the work–shopping trip chain whilst men would use the home–home evening time trip chain. The estimated marginal probability effect indicated a higher detour time sensitivity with respect to expected profit and the relative detour costs of the couriers.

*Keywords*: crowd-shipping, detour, e-groceries, mode choice, random utility theory

## 1. Introduction

Crowdsourcing has become a very popular concept in the last decade in the business and public sectors. Coined by J. Howe and M. Robinson (Howe, 2006), the crowdsourcing model was designed to outsource work to an undefined network of people – the so-called "crowd" – to involve them in activities on a volunteer or paid basis. With this concept, the business, public and healthcare sectors produced a variety of crowdsourcing-based services in public policy (Smith et al., 2015), healthcare (Weiming et al., 2016), transportation (Punel et al., 2018; Ermagun and Stathopoulos, 2018) and other fields. A further important step taken within the crowdsourcing paradigm was the involvement of occasional people in supply-related issues, such as the delivery process, which has been named crowd-shipping (CS). The concept of CS is based on the idea that commuters can act as occasional couriers and provide delivery services within their routine trips for other people. Given the different scope of the trips, CS services are available for both urban (Punel and Stathopoulos, 2017) and long-distance travel (Fly and Fetch, 2019). Given the problems of last-mile logistics, CS is first expected to address the e-shoppers' demand for urban supplies and provide a solution for deconsolidated last-mile deliveries (Buldeo Rai et al., 2018; Pourrahmani and Jaller, 2021).

CS is a new business model and it is not yet reached widely used, so evaluation of the possible impacts of CS deliveries on transport networks and urban areas remains challenging. There have been a number of studies that have examined the possible consequences of CS deliveries (Buldeo Rai et al., 2018; Chen and Chankov, 2018; Gatta et al., 2019; Dötterl et al., 2020; Simoni et al., 2020; Pourrahmani and Jaller, 2021; Tapia et al., 2023), and some have provided more comprehensive and complex evaluations utilizing agent-based behaviour (Chen and Chankov, 2018; Dötterl et al., 2020; Alho et al., 2021; Simoni et al., 2020; Tapia et al., 2023). Agent-based behaviour approaches make it possible to construct large-scale models with a high level detail in the modelled agents, and the revealed results are thus quite close to the realistic picture observed – or assumed to be so (Balmer et al., 2006).

Along with agent-based behaviour, behavioural studies allow the exploration of the possible consequences of CS deliveries by revealing mode choice patterns. The connection between consumer and courier socio-demographic characteristics and the preferred modes for CS is a key aspect that should be addressed on behavioural studies. Great attention has already been paid to the demand side of the problem, focusing on the socio-demographic characteristics of CS consumers in light of specific features of CS services (Devari et al., 2017; Punel and Stathopoulos, 2017; Punel et al., 2018; Gatta et al., 2019; Le and Ukkusuri, 2019; Fessler et al., 2022; Cebeci et al., 2023; Rossolov and Susilo, 2023). In turn, the socio-demographic characteristics of service providers have been studied with a focus on some of the specific modes used for deliveries, such as cars (Le and Ukkusuri, 2019; Ermagun et al., 2020), bikes (Wicaksono et al., 2022) and public transport (PT) (Gatta et al., 2019; Fessler et al., 2022). That said, the problem has not yet been addressed in terms of the multiple mode choice set that faces the CS courier in the transport network and what attributes she/he considers in deciding on a preferred mode of transport. To some extent, attributes such as the socio-detour (Punel and Stathopoulos, 2017) and expected remuneration for deliveries (Gatta et al., 2019; Fessler et al., 2022; Zhang et al., 2023) were studied within the behavioural framework, but they have not yet been considered jointly within the trade-off condition. The occasional couriers are also not homogenous, so it is necessary to address heterogeneity issues, starting from the perspective of gender. The delivery process is related to the physical effort needed to bear grocery packages, which justifies the necessity of exploring the gender heterogeneity of CS couriers. It can therefore be assumed that CS couriers' gender may affect behavioural aspects related to preferred transport mode with heterogeneity in terms of acceptable detours and expected remuneration amounts for males and females.

Given the research gaps identified above, this study sought to explore the mode choice behaviour of male and female occasional CS couriers concerning the trade-off conditions between acceptable detour times and expected remuneration for the delivery. Random utility maximization theory has been used as the methodological basis for revealing the heterogeneous features of the mode choice behaviour. The remainder of the paper is structured as follows: Section 2 contains a review of the literature background for mode aspects within CS deliveries. In Section 3, the study's methodology is presented, which is based on revealed preference data, simulation of travel attributes for non-chosen alternatives and formalization of mode choice probabilities. Section 4 contains the study's results, which follow from the descriptive analysis of the results for mode choice and discrete choice behaviours. The discussion of the study's findings is presented in Section 5, while Section 6 contains the concluding remarks.

## 2. Literature background

The issue of transport modes within CS deliveries has been addressed in numerous studies related to last-mile delivery aspects (Paloheimo et al., 2016; Punel and Stathopoulos, 2017; Miller et al., 2017; Arslan et al., 2019; Gatta et al., 2019; Le and Ukkusuri, 2019; Boysen et al., 2022; Fessler et al., 2022; Wicaksono et al., 2022; Tapia et al., 2023; Zhang et al., 2023). The necessity of involving occasional people in the delivery process has been considered from the prospect of remuneration to be paid, preferred detours (i.e., spatial features of CS deliveries) and their connection to the socio-demographic characteristics of the occasional couriers. This literature review section explores these issues in detail, as well as the extent to which they have been already studied.

### 2.1 Remuneration and delivery mode

The focus on public transport as the base mode for implementing eco-friendly CS deliveries can be found in studies by Gatta et al. (2019), Fessler et al. (2022) and Zhang et al. (2023). Gatta et al. (2019)

considered Rome's metro lines as the base mode for CS deliveries to promote sustainable options for e-shopping services. The willingness to act as an occasional courier was explored using random utility theory with a stated preference data collection procedure. The key feature of the study lay in accounting for the expected remuneration of €3 per delivery, which overcompensates for the trip made by Rome metro and thus promotes involvement in CS deliveries. Thus, Gatta et al. (2019) estimated the readiness of people to act as occasional couriers from the prospect of coverage of monetary expenditures for commuters ready to act as couriers.

Along with Gatta et al. (2019), Fessler et al. (2022) focused on PT mode as the basis for CS service. Using the stated preference method, behavioural data were collected to reveal the readiness of passengers from the Greater Copenhagen Area to act as CS couriers. Accounting for parcel weight, the number of parcels within one delivery and expected compensation, Fessler et al. (2022) evaluated the equilibrium conditions for people's willingness to provide the CS service. Considering a wide range of compensation for CS delivery (i.e., from 5 to 50 Danish krone), they revealed an increase in the probability of providing CS services by passengers, which indicates profit-oriented behaviour. Zhang et al. (2023) addressed remuneration aspects in a study similar to that by Fessler et al. (2022), and assessed the remuneration for CS service in a range from $1 to $5 per trip depending on the parcel size to be delivered. *Buses* were used as the base made in the simulations to estimate the reduction of carbon dioxide due to shifting parcel deliveries from conventional freight providers to PT-based CS services.

Punel and Stathopoulos (2017) considered the car mode as an appropriate option for CS services. They implemented a behavioural study based on the used the stated preference method to collect the behavioural data and evaluate the acceptable remunerations for occasional couriers for CS services. The focus was given to car-based deliveries with differentiation of the delivery cost, pickup conditions, delivery time and status of the couriers (i.e. occasional CS and professional). Covering both urban and long-distance trips, the authors estimated the mean values for remuneration as $15 and $352 for the two types of deliveries.

Using the *bike* mode for CS deliveries is essential, as it does not pollute the environment and, within the urban scale, may guarantee fast delivery speeds (Eriksson et al., 2019). Given the travel cost of zero, the expected remuneration value is an interesting aspect, as it reveals the profit-oriented behaviour of the occasional couriers. To address this issue, Wicaksono et al. (2022) proposed an equilibrium approach based on matching consumers' willingness-to-pay and couriers' willingness to provide the service. Considering a range from €1 to €13 for the cost to be paid for the delivery and the expected profit for the service provided, they defined that the balance conditions between demand and supply are guaranteed under equilibrium prices in the range from €5 to €8, depending on the delivery conditions.

Other studies that deserve to be mentioned but have not considered specific modes are Miller et al. (2017) and Le and Ukkusuri (2019), which were implemented for the United States. Both studies considered random utility maximization theory (along with Punel and Stathopoulos, 2017; Gatta et al., 2019; Fessler et al., 2022; Wicaksono et al., 2022) as a basis for evaluating the willingness to provide CS services. Considering the willingness-to-pay estimation, Miller et al. (2017) proposed willingness-to-work criteria to explore the trade-off conditions between gained profit and extra travel time for CS service. The authors revealed $19/hour as a threshold for willingness to implement the CS delivery. In turn, Le and Ukkusuri (2019) focused on trade-off conditions between travel time for CS and expected remuneration (i.e. not accounting for travel costs). Given that, the authors focused on estimating the expectation to be paid, which is similar, in a methodological sense, to willingness to pay. The explored threshold for providing the CS service estimated by Le and Ukkusuri is thus lower than that obtained by Miller et al. (2017) at $12/hour.

In summary, remuneration-related studies have significantly contributed to the understanding of couriers' willingness to provide CS services. The estimation of expected remuneration within a given range explores the shapes of the occasional couriers' behaviour from the coverage of travel costs to profit-oriented intentions. However, the connection between delivery cost, remuneration and profit expected by the occasional couriers has not yet been explored from the prospect of multiple available modes and trade-off conditions between spatial and cost-related attributes.

*2.2 Spatial and time features of detours*

The spatial features of CS detours are crucial as they can generate additional kilometres of travel. In light of the preferred mode, the detour length becomes valuable for evaluating the possible effects of CS services. Within the scope of car-based CS, Paloheimo et al. (2016) explored the impact of a CS trial project for book deliveries in the city of Jyväskylä in Finland. Based on actual case data, the experiment showed the average length of detours was 2.2 km within the spatial scope of deliveries up to 20 km. Considering the total cost optimization problem, Arslan et al. (2019) studied the time aspects of CS detours. In their optimizations, they accounted for detours in a range from 60 to 120 min with different combinations of numbers of drivers and origins of the delivery. Minimizing the cost function, the authors considered the CS detour as an element to improve the total cost; in this case, the CS should ensure reduced vehicle kilometres travelled. The store origin for CS trips was defined as most relevant to achieving the goals concerning the detours in the studied range.

In turn, Boysen et al. (2022) evaluated the possibility of using distribution centres and sales outlets as origin points for CS deliveries with the involvement of their employees in CS services. They considered the problem of matching demand and supply systems to provide efficient and environmentally friendly CS-based deliveries. Private cars were taken as the base mode within the study with a focus on such delivery aspects as implemented detours and expected remuneration for the CS delivery. Although Arslan et al. (2019) and Boysen et al. (2022) focused on the optimization aspects of CS deliveries, the detour-related findings are a good foundation for formalizing the mode choice problem within the CS field.

Tapia et al. (2023) developed an agent-based model to evaluate the effects of the introduction of CS deliveries. The activity of every agent (CS courier) in the model was determined by such attributes as CS detours and its connection to the remuneration expected. Two travel modes were considered, *car* and *bike*, which allowed the authors to reveal the spatial and environmental-related issues of CS delivery services. They revealed crucial aspects of CS deliveries in the light of the detour aspect, distinguishing them between *car* and *bike* modes. The simulations results showed that the average detour distance covered by car-based CS couriers is about 10 km per parcel, while the CS detour is slightly lower for *bike*-based trips at 9 km.

Summarizing the spatial aspects of CS detours, it should be emphasized that the behavioural aspects of the CS couriers should be considered. Accounting for the random utility maximization theory, occasional couriers are expected to act from the prospect of their personal utility while evaluating the acceptable delivery distances and times for CS services. The insights on the interconnection between acceptable detour lengths and preferred mode for CS delivery have also not yet been explored and will be covered within this study.

*2.3 Socio-demographic characteristics of couriers*

Evaluation of socio-demographic characteristics is an essential procedure within most studies analysed in this section. For instance, Gatta et al. (2019) focused on the age of potential couriers, while Le and Ukkusuri (2019) explored education level, race and gender, along with age characteristics. Fessler et al. (2022) focused on assessing the age and employment status of PT users ready to provide CS services. In most cases, socio-demographic characteristics are incorporated into discrete choice models to account for the observed heterogeneity of choices (Bhat, 2000). This approach provides researchers with insight into perceived utilities by the decision-makers and trace the connections between their socio-demographic characteristics and the choices made. On other hand, the baseline conditions, which should be taken into account, require omitting some of the socio-demographic characteristics in the normalization procedure (Train, 2009). For instance, the gender aspect is usually presented as a binary component, with the selection of "male" or "female" as the base during the behaviour. In this case, the behaviour results represent the decision-maker's attitude with respect to the chosen base socio-demographic characteristic, while the non-chosen attribute is considered as, *per se* following the baseline characteristics but with the opposite meaning. Given that, this study explored the gender heterogeneity of occasional couriers using both random utility maximization theory and descriptive analysis techniques. Special attention has been given to gender aspects in their relation to acceptable detours and remuneration in the context of the preferred modes for CS deliveries.

## 3. Methodology

### 3.1 Behaviour framework

Considering the problem of mode choice for new delivery services, capturing people's actual experience within the transport network is essential. In the case of CS services, occasional couriers should make CS deliveries within their daily routine trips with some detours that are acceptable to the occasional couriers. Given that, the problem considered combines the existing travel patterns of the occasional couriers with the new spatial features of the trips to be fulfilled.

Due to specific features of the different modes of transport (e.g. travel speeds, travel costs and level of comfort), the detours acceptable to occasional couriers are expected to differ. Having the travel patterns of people who are ready to act as occasional CS couriers, the preferred mode for CS deliveries can be explored based on the revealed preferences for routine trips. Following the methodology for assessing mode choice for regular trips, the proposed methodological framework collects behavioural data on acceptable detour times, preferred modes for CS deliveries and expected remuneration values to form the dataset for a chosen alternative. The spatial and time characteristics of non-chosen alternatives are also reconstructed using the PTV Visum simulation environment. The researcher group can thus form the dataset needed to implement the descriptive and discrete choice analyses for exploring the mode choice behaviour of occasional CS couriers. Figure 1 depicts the proposed methodological framework of the study.

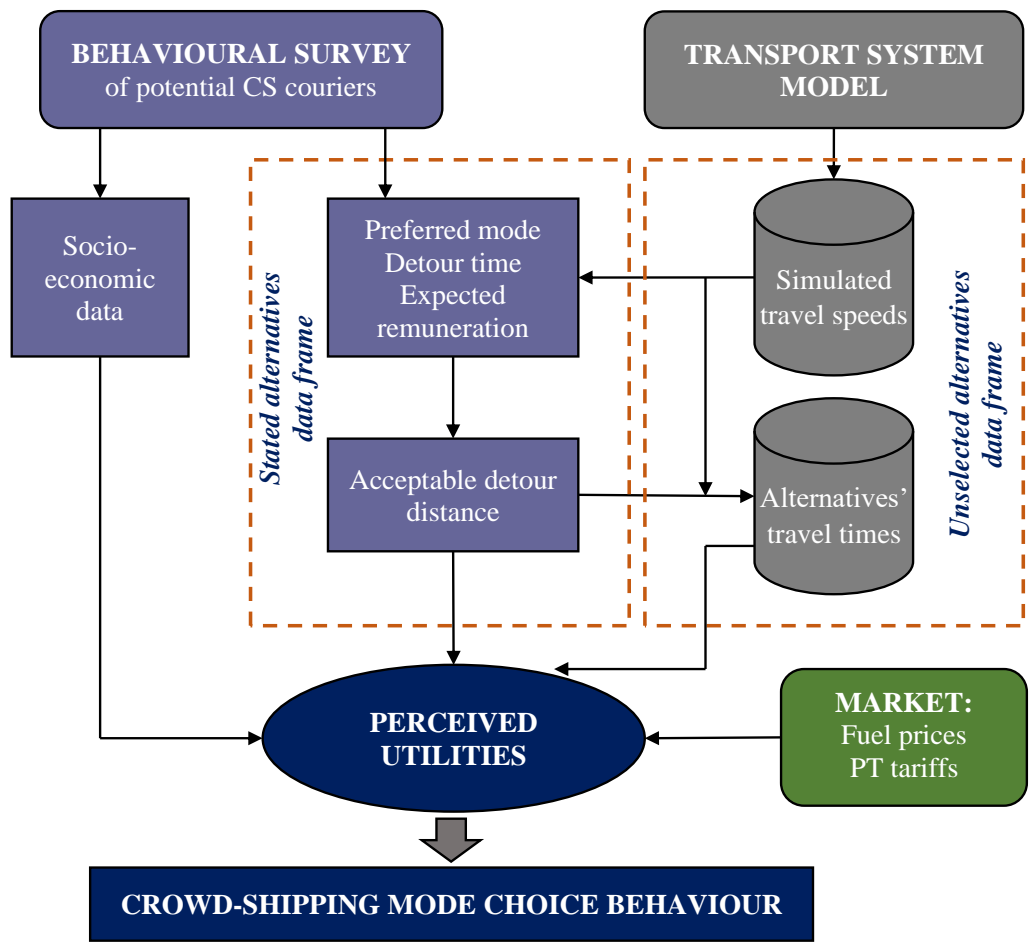

**Fig. 1. Methodological framework for estimation CS mode choice behaviour**

The initial data for assessing the mode choice behaviour in the context of the CS service were collected through a behavioural survey. As the current travel patterns matter, the survey aimed to reveal the respondents' attitude towards the following:

- preferred mode for implementation of CS delivery;
- detour time acceptable to the potential courier in the context of the revealed preferred mode;
- detour type for CS delivery (i.e., its integration into routine trip pairs); and
- expected remuneration to be paid for a CS delivery service in the frame of the acceptable detour time.

Figure 2 presents the example of possible detour types, as well as the concept of CS deliveries for e-groceries.

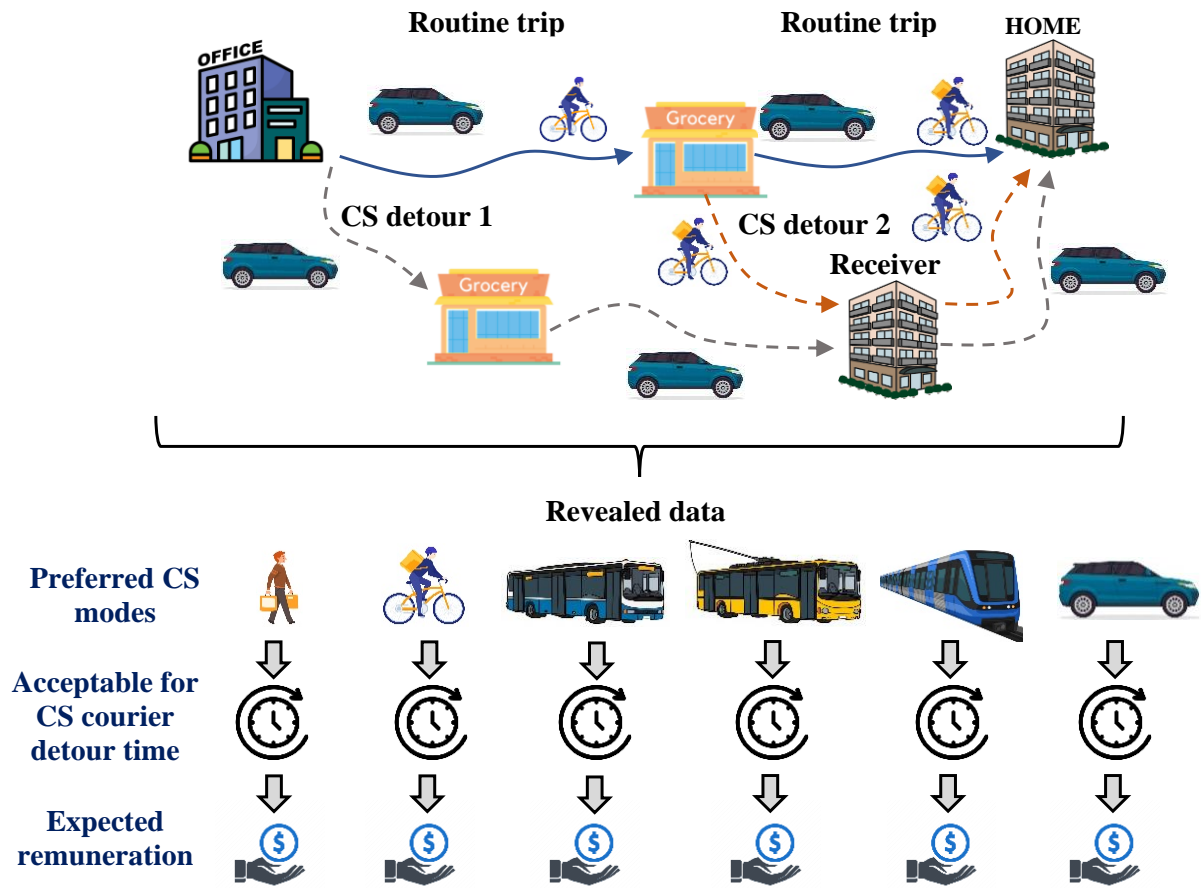

**Fig. 2. Instance of behavioural data collection framework**

Once the preferred mode was revealed, the procedure for data reconstruction of existing alternative modes was followed. To reach that goal, the following simulation procedure is proposed. PTV Visum software provides the possibility to simulate the travel-related attributes within a multimodal transport system. Despite some limitations as a zone-based simulation, PTV Visum makes it possible to obtain the dataset needed for evaluating the spatial and time features of the trips made. The zoning of the simulated area is crucial in this case, and the modeler should pay special attention to this issue. The general description of the combination of the revealed and simulated data is presented in Figure 1. To provide a more detailed description of this process, Figure 3 depicts an instance of the data formation flowchart based on the revealed and simulated dataset for one CS delivery.

According to Figure 3, information channel #1 provides the data on the detour time acceptable to the CS courier (i.e. the time threshold for readiness to provide the service). Given the information on the preferred travel mode, the evaluation of the detour distance is made via channel #2 and resulted in the database via channel #3. According to the estimated detour distance based on the preferred travel mode, the detour times for the non-chosen alternative modes are estimated (flow #4 interaction with PTV Visum behaviour environment). Figure 3 presents the results of this process for information flow #5. Following this process results in the formation of a database of spatial and time attributes.

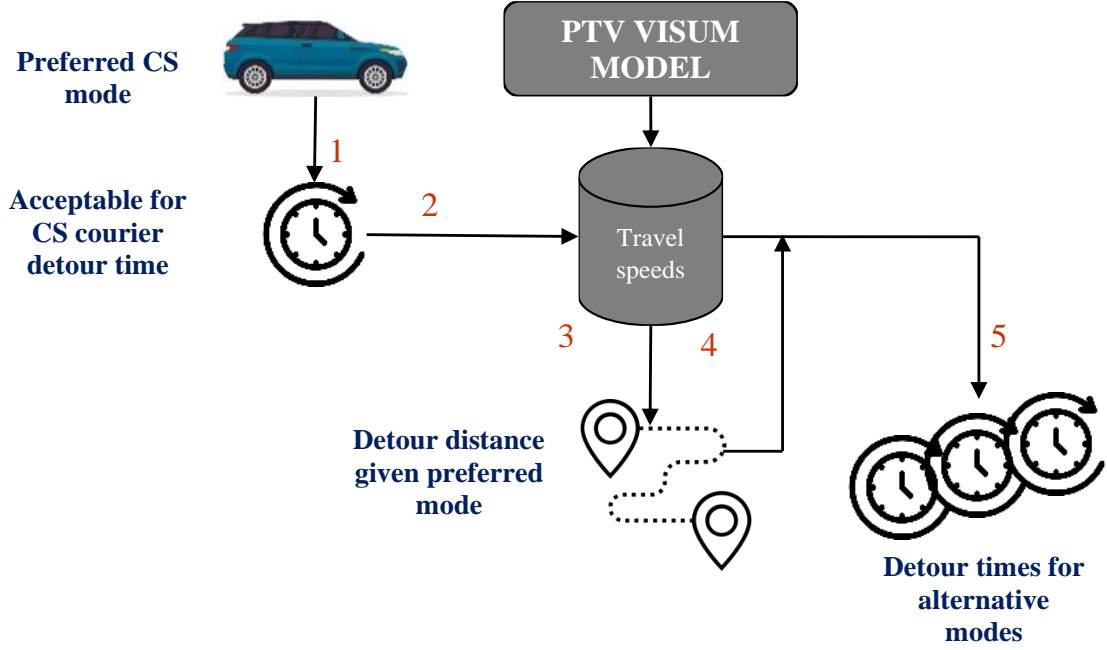

**Fig. 3. Instance of the data formation flowchart for time and spatial attributes**

The proposed methodological framework allows the collection of the behavioural data needed for the estimation of the mode choice attitudes for CS deliveries. The next section contains the description of developed methodology for exploring CS mode choice behaviour.

### 3.2 CS mode choice model

Having a set of alternative modes $\Omega = \{m_k\}, k \in \mathbb{N}, k \in M$ where every mode can be presented by quality $t_k$ and cost $f_k$ attributes (i.e. $m_k = \langle t_k, f_k \rangle, t_k > 0, f_k > 0$) the transport network user $i$ makes the choice of the mode for travel $q$ through the evaluation of the context set $\Omega_{qi} = \{\langle t_{qk}, f_{qk} \rangle\}, k \in \mathbb{N}, k \in M$. Following the rule that commuters tend to reduce time and cost expenditures, it is expected that every traveller wants to increase the perceived utility by trading off time and cost. Given that, every subset $m_k$ of alternative modes $M$ for travel $q$ can be represented by the utility $U_{kqi}$. In this context, commuter $i$ chooses mode $k$ for travel $q$ under the following condition $U_{kqi} = \max_{k \in M} \forall U_{kqi}$ (Ben-Akiva and Lerman, 1985; Hensher et al., 2005; Train, 2009; Ortuzar and Willumsen, 2011). Following the rule that "only the difference in the utilities matters" (Train, 2009), the probability of $k$ mode being chosen within $M$ alternatives is as follows:

$$
\begin{aligned}
P_{kqi} &= \text{Prob}\left\{ V_{kqi} + \varepsilon_{kqi} > V_{dqi} + \varepsilon_{dqi}, k, d \in M \right\} \\
&= \text{Prob}\left\{ V_{kqi} - V_{dqi} > \varepsilon_{dqi} - \varepsilon_{kqi}, k, d \in M \right\} \\
&= \text{Prob}\left\{ \begin{array}{l} \left( \beta_{t(k)} \cdot t_{kqi} + \beta_{f(k)} \cdot f_{kqi} \right) - \left( \beta_{t(d)} \cdot t_{dqi} + \beta_{f(d)} \cdot f_{dqi} \right) > \\ \varepsilon_{dqi} - \varepsilon_{kqi}; k, d \in M; \beta_{t(k,d)} \neq \beta_{f(k,d)} \end{array} \right\}
\end{aligned} \tag{1}
$$

where $V_{kqi}$ is deterministic part of the utility function, $\varepsilon_{kqi}$ is the independent and identically distributed extreme value, $\beta$ is the coefficient of the model and $M$ is a number of alternative modes considered.

*Eq.* 1 represents the equilibrium for mode choice when travel times (quality of service) and travel costs are under trade-off conditions. In the context of the CS service, the occasional courier receives a remuneration of value $r$ for the CS service. As a trade-off to that, the courier has to make a detour that results in additional time expenditures, as $\Delta t_{kqi}$. It should be noted that the respondent states the value of the preferred remuneration $r$, which is why $r$ does not vary across the alternatives $M$ for trip $q$. This statement does not apply to $\Delta t_{kqi}$. This is justified by the fact that the value of $\Delta t_{kqi}$ is stated by the respondent for the preferred mode, while for other modes it should be estimated based on the simulations. In these conditions, it is obvious that the time for reaching the destination point using other modes will be not the same (i.e. $\Delta t_{kqi} \neq \Delta t_{dqi}$). That said, for the CS delivery, *Eq.* 1 should be enhanced with remuneration and detour-related attributes, which results in the following:

$$P_{kqi}^{CS} = \text{Prob} \left\{ \begin{array}{l} \left( \beta_{t(k)} \cdot \underbrace{(t_{kqi} + \Delta t_{kqi})}_{\text{total travel time for } k} + \beta_{f(k)} \cdot \underbrace{(f_{kqi} + \Delta f_{kqi})}_{\text{total travel cost for } k} + \beta_r \cdot r_{qi} \right) - \\[2em] \left( \beta_{t(d)} \cdot \underbrace{(t_{dqi} + \Delta t_{dqi})}_{\text{total travel time for } d} + \beta_{f(d)} \cdot \underbrace{(f_{dqi} + \Delta f_{dqi})}_{\text{total travel cost for } d} + \beta_r \cdot r_{qi} \right) > \varepsilon_{dqi} - \varepsilon_{kqi}; k, d \in M; \beta_{t(k,d)} \neq \beta_{f(k,d)} \end{array} \right\} \quad (2)$$

*Eq.* 2 accounts for the total travel times and costs that reflect the full trip chain implemented by traveller $i$ during the day. In framework of the CS delivery, the components $t_{qi}$ and $f_{qi}$ do not influence the difference between the utilities, thus acting as constants. *Eq.* 2 can therefore be simplified by exclusion these components. Another issue that should be addressed is the added remuneration value $r_{qi}$, which should be normalized for one alternative to zero, as it does not vary across the alternatives (Train, 2009). Once all of these procedures are implemented, *eq.* 2 transforms into the following:

$$P_{kqi}^{CS} = \text{Prob} \left\{ \begin{array}{l} \left( \beta_{t(k)} \cdot \Delta t_{kqi} + \beta_{f(k)} \cdot \Delta f_{kqi} + \beta_r \cdot r_{qi} \right) - \\[1em] \left( \beta_{t(d)} \cdot \Delta t_{dqi} + \beta_{f(d)} \cdot \Delta f_{dqi} \right) > \varepsilon_{dqi} - \varepsilon_{kqi}; k, d \in M; \beta_{t(k,d)} \neq \beta_{f(k,d)} \end{array} \right\} \quad (3)$$

*Eq.* 3 describes the possibility of mode choice between $M$ alternatives based on the detour made that is acceptable to the CS occasional courier. The proposed theoretical frame provides the possibility for evaluating the profit-oriented behaviour of occasional couriers. CS is known as a society-oriented service, which means that people are ready to contribute through their activities to the community's welfare – for instance, in terms of external cost reduction. Once CS service is deployed, however, people may consider it as another business model for monetary gain, in which case the expected remuneration is key attribute in the behaviour framework, as it makes it possible to trace the profit-oriented behaviour of occasional couriers. The mathematical interpretation of mode choice accounting for the profit maximization attitude is as follows:

$$P_{kqi}^{CS} = \text{Prob} \left\{ \begin{array}{l} \left( \beta_{t(k)} \cdot \Delta t_{kqi} + \beta_{w(k)} \cdot \underbrace{(r_{qi} - \Delta f_{kqi})}_{\text{profit for } k} \right) - \left( \beta_{t(d)} \cdot \Delta t_{dqi} + \beta_{w(d)} \cdot \underbrace{(r_{qi} - \Delta f_{dqi})}_{\text{profit for } d} \right) > \\[2em] \varepsilon_{dqi} - \varepsilon_{kqi} \\[1em] = \left( \beta_{t(k)} \cdot \Delta t_{kqi} + \beta_{w(k)} \cdot w_{kqi} \right) - \left( \beta_{t(d)} \cdot \Delta t_{dqi} + \beta_{w(d)} \cdot w_{dqi} \right) > \varepsilon_{dqi} - \varepsilon_{kqi}; \\[1em] k, d \in M; \beta_{t(k,d)} \neq \beta_{w(k,d)} \end{array} \right\} \quad (4)$$

In such conditions, modes free of travel cost should be revealed with dominant patterns of their usage. This statement stands, in particular, for the *bike* mode, as it has high travel speeds (Eriksson et al., 2019), similar to the *car* mode, but without the travel cost – that is, it could guarantee the maximum profit for the occasional courier. Using *Eq.* 4, it is easy to demonstrate these dominant conditions by, for example, comparing the pair of *bike* and *bus* modes. Considering this instance, the utility for the *bike* mode will be higher given the following inequalities: $\Delta t_{bike} < \Delta t_{bus} \wedge (r - 0_{bike}) > (r - \Delta f_{bus})$. To capture the random distribution effect for $\beta$, we propose using the mixture of the logit model (Train, 2009):

$$P_{kqi}^{CS} = \int L_{kqi} \cdot f(\beta) \cdot d\beta = \int \left( \frac{e^{\sum_{j \in N} \beta_{jk} \cdot x_{jkqi}}}{\sum_{k \in M} e^{\sum_{j \in N} \beta_{jk} \cdot x_{jkqi}}} \right) \cdot \phi(\beta \,|\, b, W) \cdot d\beta \tag{5}$$

where $L_{kqi}$ is the multinomial logit function, $f(\beta)$ is the density of continuous distribution, $b$ and $W$ are the mean and variance of the continuous distribution, respectively, and $N$ is the number of choice attributes accounted for in the deterministic part of the utility function.

## 4. Results

### 4.1 Survey and sample descriptions

The behavioural data were collected online in February–March 2021 in Kharkiv, Ukraine. The survey covered demand and supply-related issues to explore the readiness of people to use CS and act as occasional CS couriers. The administration and dissemination of the questionnaire was done by the Ukrainian part of the research team. The snowball sampling technique was used, as it allows researchers to cover the relevant sample under the constraint of a low budget (Biernacki and Waldorf, 1981).

The demand-side for CS deliveries and the expected level of the services were revealed based on the stated-preference method, with further development of a hybrid choice framework. The data needed for such evaluation were collected within second and third sections of the questionnaire (Annex A). The results on the demand side from the CS survey are presented in Rossolov and Susilo (2023). The supply side of the CS deliveries was covered in the fourth part of the questionnaire, in which respondents were asked to provide information about acceptable detour times if they acted as the CS couriers, as well as the mode to be used for CS service. The survey participants were also asked to describe the preferred time of day when they were ready to implement CS delivery. As CS service is expected to take place within the routine trip chains fulfilled by commuters, the question of how the CS delivery process would be integrated into people's trip chains was addressed in the questionnaire to allow the research team to reveal the mode choice behaviour in more detail. To account for possible profit-oriented behaviour, the information on the expected value of remuneration for CS delivery by occasional couriers was collected.

The survey covered 287 persons, and after data cleaning, the final sample was 249 respondents. The breakdown of the sample is presented in Table 1. As shown in Table 2, the sample formed is well balanced in terms of gender and is very close to representative of the total population. As the research team aimed, within this survey, to reveal the mode choice behaviour for CS deliveries accounting for the gender-related issue, the formed sample is considered appropriate for the purpose. The car availability information also is important in light of the topic studied. The prevailing number of car owners (66.67%) in the sample provides the possibility for evaluating the environmental issues if *car*-oriented behaviour for CS deliveries is revealed.

**Table 1. Breakdown of the sample**

| Attribute | Frequency | Sample, % | Population, % |
|---|---|---|---|
| Age | | | |
| 18–24 | 81 | 32.53% | 14.00% |
| 25–34 | 58 | 23.30% | 18.00% |
| 35–44 | 62 | 24.90% | 19.00% |
| ≥ 45 | 48 | 19.28% | 49.00% |
| Gender | | | |
| Female | 121 | 48.59% | 53.64% |
| Male | 128 | 51.41% | 46.36% |
| Household size | | | |
| 1 person | 55 | 22.09% | 24.90% |
| 2 persons | 50 | 20.08% | 33.60% |
| 3 persons | 78 | 31.33% | 26.80% |
| 4 persons and more | 48 | 26.51% | 14.70% |
| Car availability in household | | | |
| Yes | 166 | 66.67% | NA* |
| No | 83 | 33.33% | |
| Monthly personal wage, UAH** | | | |
| < 5,000 | 28 | 11.24% | 11.80% |
| 5,000–9,999 | 83 | 33.33% | 19.70% |
| 10,000–19,999 | 80 | 32.13% | 35.90% |
| 20,000–29,999 | 30 | 12.05% | 23.10% |
| 30,000–39,999 | 6 | 2.41% | 5.30% |
| 40,000–49,999 | 9 | 3.62% | 2.60% |
| ≥ 50,000 | 13 | 5.22% | 1.60% |
| Employment status | | | |
| Full-time | 155 | 62.25% | |
| Part-time | 52 | 20.88% | |
| Unemployed | 3 | 1.21% | NA |
| Housekeeper | 15 | 6.02% | |
| Student | 24 | 9.64% | |

Note:  *NA – Not Available
  **Currency exchange at the times of the survey (February–March 2021) was 27.9 UAH to $1.

*4.2 Transport network and simulation environment*

The transport network of the city of Kharkiv at the time of the survey consisted of four modes within PT segments: *bus*, *metro*, *trolleybus* and *tram*. All of these modes have different transport operators, which imposes some peculiarities on the tariff systems provided to the commuters. Kharkiv covers 350 km² and the total length of the roads of different types is 1950.97 km. Figure 4a provides a detailed presentation of the spatial distribution of roads within the city's area by type. The basic numerical characteristics of PT modes are presented in Table 2.

**Table 2. Basic characteristics of the PT network**

| Mode | Number of lines, units | Total length of lines, km | Tariff, UAH |
|---|---|---|---|
| Bus | 97 | 924.55 | 10 |
| Metro | 3 | 38.60 | 8 |
| Trolleybus | 33 | 262.29 | 6 |
| Tram | 13 | 166.29 | 6 |

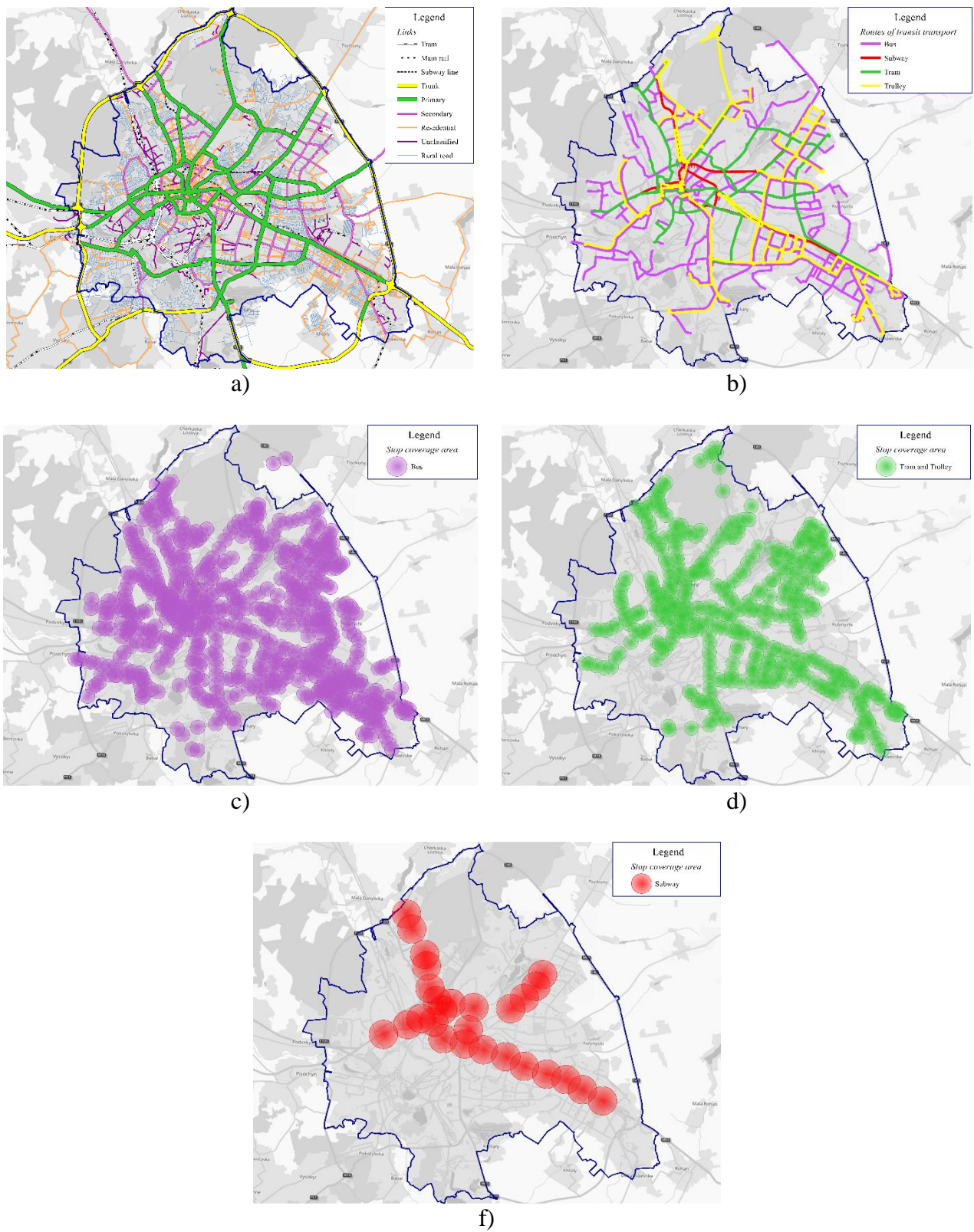

**Fig. 4. Basic characteristics of the transport system of Kharkiv, represented in PTV Visum model:**
a) road network; b) public transport network; c) spatial coverage of bus stops; d) spatial coverage of tram and trolleybus stops; f) spatial coverage of metro stations

Based on the information presented in Figure 4 and Table 3, it can be concluded that the road and PT networks of Kharkiv are well developed and provide a good opportunity for commuters to choose the preferable mode for their trips within the urban scale. Looking at CS deliveries, such conditions provide relatively high flexibility for occasional couriers regarding the implementation of detours and use of the same mode for servicing e-shoppers. Given that *trolleybus* and *tram* modes have equal tariffs and similar operational features, they have been combined within the survey into one mode called *electric ground PT*. In the figures, this mode is presented by the image of a trolleybus.

### *4.3 Descriptive analysis*

Acceptable detour times were determined for all 249 survey participants. This information was structured according to the preferred modes to connect it further with the time and cost-related attributes of CS deliveries. The mode choice data were handled to account for gender heterogeneity. Figure 5 presents the mode choice preferences accounting for the gender information of the occasional CS couriers.

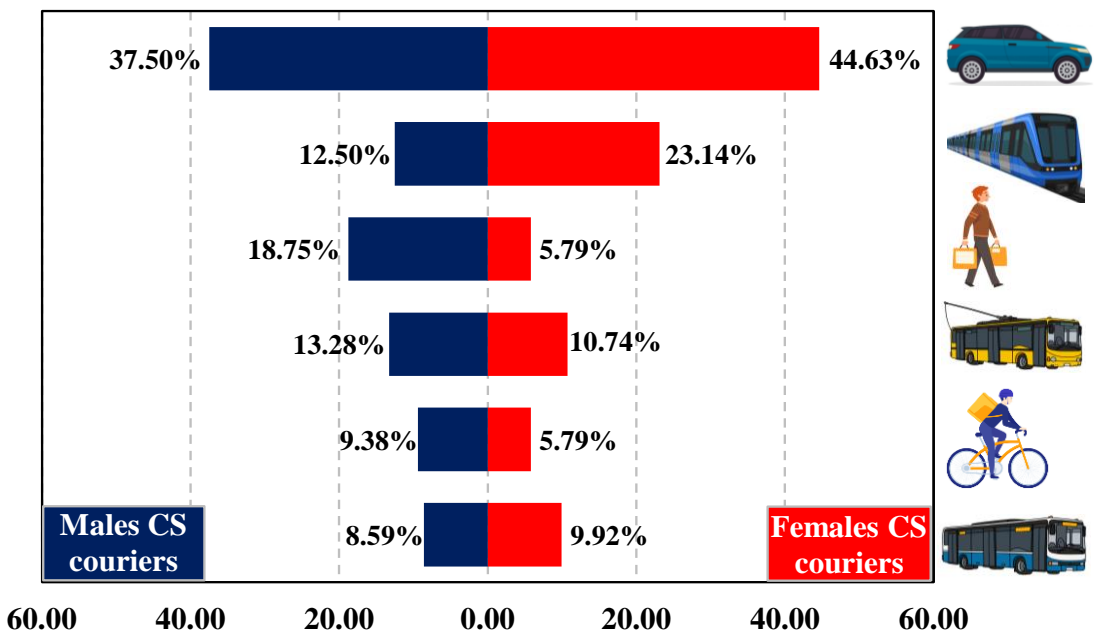

**Fig. 5. Revealed mode choice preferences for CS deliveries by occasional couriers**

The preliminary analysis of the mode choice behaviour allowed the research group to identify the *car*-oriented behaviour for both genders, but with slight predominance for women. In turn, women were more likely to use the *metro* for CS – that is, 23.14% of the female participants were likely to use the *metro*, compared to 12.50% of male respondents. The opposite behaviour was revealed for the *walking* mode, as 18.75% of men were ready to implement CS delivery using just their own physical capabilities, while only 5.79% of women were ready to do so, and nearly identical behaviour was detected for the *bike* mode, where 9.38% men would prefer to use this mode, while only 5.79% of women shared this attitude. For *electric ground transport* and *buses*, the mode choice behaviour of men and women were quite similar. The revealed *car*-oriented demeanour may produce a high level of environmental pollution once CS service reaches wide usage among e-shoppers, in which case, the detour times for implementing the CS delivery that are acceptable to occasional couriers become highly valuable for evaluating the possible externalities of CS deliveries.

The acceptable detour times and the preferred modes are appropriate information for the expected impact of CS deliveries on the city area. The acceptable detour times proposed for evaluation by the respondents were in the range of 15 to 60 minutes; the descriptive statistics are summarized in Table 3.

**Table 3. Descriptive statistics on acceptable detour times (in minutes)**

| Mode | Mean | | Standard deviation | | Minimum | | Maximum | |
|---|---|---|---|---|---|---|---|---|
| | Female | Male | Female | Male | Female | Male | Female | Male |
| Bike | 27.9 | 35.0 | 10.35 | 9.77 | 15 | 15 | 45 | 45 |
| Bus | 30.0 | 15.0 | 18.09 | 0.00 | 15 | 15 | 60 | 15 |
| Car | 41.1 | 38.4 | 16.01 | 16.64 | 15 | 15 | 60 | 60 |
| Metro | 37.5 | 31.9 | 13.23 | 13.28 | 15 | 15 | 60 | 45 |
| Walking | 30.0 | 42.5 | 17.32 | 17.51 | 15 | 15 | 60 | 60 |
| Electric ground PT | 38.1 | 47.6 | 16.90 | 12.13 | 15 | 15 | 60 | 60 |

According to the mean values for the detour times, women appeared to be ready to implement slightly longer detours by *car* and *metro* compared to men. In turn, if active modes (*bike* and *walking*) are used for CS deliveries, men were ready to make longer trips than women, and the spread between the average detour times was higher than was detected for the *car* and *metro* modes. The *bus* mode was revealed as not attractive for male respondents, as they expressed their readiness to use this delivery option only for short detours (i.e. 15 minutes), which confirms the information about preferred transport modes (Fig. 5). Women also engaged the *bus* mode with non-solid behaviour, as the standard deviation for the detour time was the highest for this mode in the considered list. In the frame of the maximum acceptable detour times, men and women revealed the same behaviour only for *bike* mode, with a limit of up to 45 minutes. While men were ready to use the *metro* mode for no more than 45 minutes for CS detours.

The revealed acceptable detour times and preferred mode information needs to be enhanced with the expected remuneration for CS deliveries, as this allows detection of profit-oriented behaviour. Respondents were therefore asked to choose a preferable remuneration amount in the range of 60 and 120 UAH. These values were taken from the SP experiments to identify the match between demand and supply subsystems for the delivery cost and remuneration. Figure 6 summarizes these data in the form of box and whisker charts.

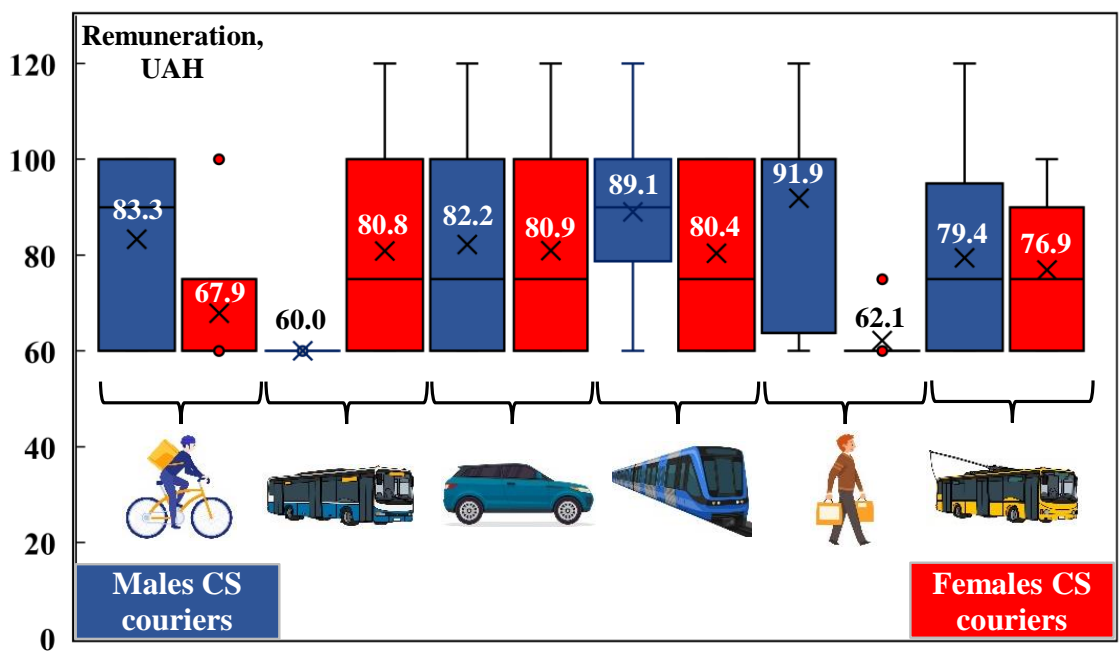

**Fig. 6. Box and whisker charts for expected remuneration for CS delivery services**

The evaluation of the results revealed that women expected 8% less remuneration than men within the considered modes. The *bike* and *walking* modes had the highest differences in remuneration, with 18.5% and 32.4%, respectively, less for women couriers. Given that active modes do not have any travel cost, the preliminary conclusion about this effect can be summarized as men having a greater profit orientation than women. The expected remuneration values also represent the perceived value of time. The highest mean value for remuneration was, for example, determined for the *walking* mode for male CS couriers. For the *bike* and *metro* modes, men also had relatively high expected remuneration values. The necessity for sufficient physical activity can explain such phenomena if these modes are used for CS service. For the *metro* mode, physical activity is involved in the long walking distance to reach the stations (see Fig. 4). For women, active modes were already revealed as less preferable (see Fig. 5), which is why the obtained values for the expected remuneration are not particularly high. Such findings also confirm the assumption that women perceive the value of their time to be lower than men do.

In light of these findings, the integration of the detour into the daily activities of the occasional couriers is a highly valuable aspect of the service. The research team also aimed to investigate the frequency of deliveries that CS couriers were ready to provide, and these issues were covered during the behavioural survey. The obtained descriptive statistics on the preferred integration of detours into daily trips and the frequency of deliveries planned are summarized in Tables 4 and 5.

**Table 4. Breakdown of preferred integration of CS detours into routine trips**

| Specification | Preferred trip-chain integration | | | | | | | |
|---|---|---|---|---|---|---|---|---|
| | **H-W** | | **W-H** | | **W-G** | | **H-H\*** | |
| | Units | % | Units | % | Units | % | Units | % |
| *Within the sample:* | | | | | | | | |
| Male | 12 | 9.38 | 47 | 36.72 | 20 | 15.63 | 49 | 38.28 |
| Female | 20 | 16.53 | 47 | 38.84 | 30 | 24.49 | 24 | 19.83 |
| *Mode-based evaluation:* | | | | | | | | |
| **Bike** | | | | | | | | |
| Male | 1 | 8.33 | 9 | 75.00 | 0 | 0.00 | 2 | 16.67 |
| Female | 4 | 57.14 | 2 | 28.57 | 0 | 0.00 | 1 | 14.29 |
| **Bus** | | | | | | | | |
| Male | 0 | 0.00 | 0 | 0.00 | 0 | 0.00 | 11 | 100.00 |
| Female | 1 | 8.33 | 1 | 8.33 | 7 | 58.33 | 3 | 25.00 |
| **Car** | | | | | | | | |
| Male | 4 | 8.33 | 20 | 41.67 | 7 | 14.58 | 17 | 35.42 |
| Female | 8 | 14.81 | 28 | 51.85 | 11 | 20.37 | 7 | 12.96 |
| **Metro** | | | | | | | | |
| Male | 3 | 18.75 | 7 | 43.75 | 3 | 18.75 | 3 | 18.75 |
| Female | 2 | 7.14 | 13 | 46.43 | 6 | 21.43 | 7 | 25.00 |
| **Walking** | | | | | | | | |
| Male | 4 | 16.67 | 9 | 37.50 | 2 | 8.33 | 9 | 37.50 |
| Female | 0 | 0.00 | 1 | 14.29 | 4 | 57.14 | 2 | 28.57 |
| **Electric ground PT** | | | | | | | | |
| Male | 0 | 0.00 | 2 | 11.76 | 8 | 47.06 | 7 | 41.18 |
| Female | 5 | 38.46 | 2 | 15.38 | 2 | 15.38 | 4 | 30.77 |

Note: H-W – "Home–Work"; W-H – "Work–Home"; W-G – "Work–Groceries"; H-H – "Home–Home (evening)"

**Table 5. Breakdown of preferred frequency of deliveries to be provided**

| Specification | Everyday | | Several times a week | | Once a week | | Several times a month | | Once a month | | Less than once a month | |
|---|---|---|---|---|---|---|---|---|---|---|---|---|
| | Units | % | Units | % | Units | % | Units | % | Units | % | Units | % |
| *Within the sample:* | | | | | | | | | | | | |
| Male | 5 | 3.90 | 25 | 19.53 | 40 | 31.25 | 16 | 12.50 | 21 | 16.41 | 21 | 16.41 |
| Female | 6 | 4.96 | 33 | 27.27 | 25 | 20.66 | 15 | 12.40 | 22 | 18.18 | 20 | 16.53 |
| *Mode-based evaluation:* | | | | | | | | | | | | |
| **Bike** | | | | | | | | | | | | |
| Male | 0 | 0.00 | 2 | 16.67 | 2 | 16.67 | 1 | 8.33 | 4 | 33.33 | 3 | 25.00 |
| Female | 1 | 14.29 | 0 | 0.00 | 3 | 42.86 | 1 | 14.29 | 2 | 28.56 | 0 | 0.00 |
| **Bus** | | | | | | | | | | | | |
| Male | 0 | 0.00 | 6 | 54.55 | 0 | 0.00 | 0 | 0.00 | 0 | 0.00 | 5 | 45.45 |
| Female | 0 | 0.00 | 4 | 33.33 | 0 | 0.00 | 0 | 0.00 | 2 | 16.67 | 6 | 50.00 |
| **Car** | | | | | | | | | | | | |
| Male | 1 | 2.09 | 9 | 18.75 | 16 | 33.33 | 7 | 14.58 | 7 | 14.58 | 8 | 16.67 |
| Female | 1 | 1.85 | 22 | 40.74 | 5 | 9.26 | 6 | 11.11 | 8 | 14.81 | 12 | 22.23 |
| **Metro** | | | | | | | | | | | | |
| Male | 0 | 0.00 | 6 | 37.5 | 4 | 25.00 | 2 | 12.50 | 3 | 18.75 | 1 | 6.25 |
| Female | 4 | 14.29 | 6 | 21.43 | 8 | 28.57 | 5 | 17.86 | 4 | 14.29 | 1 | 3.56 |
| **Walking** | | | | | | | | | | | | |
| Male | 3 | 12.50 | 1 | 4.17 | 10 | 41.66 | 1 | 4.17 | 6 | 25.00 | 3 | 12.50 |
| Female | 0 | 0.00 | 0 | 0.00 | 3 | 42.86 | 2 | 28.56 | 1 | 14.29 | 1 | 14.29 |
| **Electric ground PT** | | | | | | | | | | | | |
| Male | 1 | 5.88 | 1 | 5.88 | 8 | 47.06 | 5 | 29.42 | 1 | 5.88 | 1 | 5.88 |
| Female | 0 | 0.00 | 1 | 7.69 | 6 | 46.15 | 1 | 7.69 | 5 | 38.47 | 0 | 0.00 |

The analysis of the CS detour integration into trip pairs allowed the research group to identify the Work–Home and Home–Home (evening time) trip pairs as preferable for CS-related detours. Such behaviour is more pronounced for male CS couriers, as the mentioned trip pairs have almost equal shares in the sample. Particular attention should be paid to the Home–Home (evening time) trip pair, which prevails for the male sample share – that is, 38.28% of men chose this trip chain integration. CS deliveries would, in this case, generate additional kilometres travelled, providing additional pressure on the transport system and network. Considering preferred modes, this aspect is crucial, as the Home–Home (evening time) trip pair has the highest share for the *car* mode. The *bus* mode is also revealed as preferable for the Home–Home (evening time) trip pair. Still, having very low acceptable detour times (see Table 3), the production of extra trips on the *bus* due to CS deliveries is less sensitive than the *car* mode in terms of external costs.

Female respondents expressed detour behaviours with an emphasis on the Work–Home (38.84%) and Work–Groceries (24.49%) trip pairs. Female CS couriers are thus expected to create lower pressure on the transportation network than men. Both preferred types of CS detour integration are less harmful than Home–Home (evening time), which was found to be the favourable trip pair integration for male couriers. On the other hand, women CS couriers showed more *car*-oriented behaviour, with higher acceptable CS detour times than men (see Table 3). Moreover, the frequency of the CS deliveries that women were ready to provide using a *car* mode is higher than for male CS couriers (see Table 5). For instance, the delivery option "several time a week" was preferable for *car*-based women CS couriers, while for men, the option "once a week" prevailed. Comparing the general distribution of the responses within the entire range of delivery frequency, women's behaviour tended to have uniform distribution across all levels. The responses of male CS couriers showed a peak for the once a week frequency.

The provided descriptive statistics analysis should be enhanced with discrete choice modelling results to explore the mode choice behaviour for CS deliveries by occasional couriers. Once the results have been obtained, the summary of the preferred modes for CS services can be formalized and the possible effects of CS deployment depicted.

*4.4 Mode choice modelling results*

The simulation of travel speeds is an initial step in forming the necessary dataset. This procedure was implemented in PTV Visum. The results obtained for all modes considered in this study are summarized in Table 6. The simulations were made considering the distribution of PT stops and stations within the city's transport network, which allowed the research group to account for the walking distances involved in access and egress. The dwell time at PT stops was also considered. The *tram* and *trolleybus* modes were again combined into *electric ground PT*, as they have essentially equal travel speeds and equal tariffs.

**Table 6. Initial data for estimation travel times and costs**

| Mode | Simulated travel speeds, km/h | Source for calculating travel cost | |
|------|---|---|---|
| | | Tariff per trip, UAH | Price per 1 litre of fuel, UAH/1liter |
| Bike | 25 | — | — |
| Bus | 9 | 10 | — |
| Car | 23 | — | 27 |
| Metro | 13 | 8 | — |
| Walking | 4 | — | — |
| Electric ground PT | 7.5 | 6 | — |

In the discrete choice modelling, the research group aimed to cover economic-related issues that were not addressed within the descriptive analysis. The focus was on revealing the connections between income attributes and mode choice for CS deliveries. The *walking* mode was referred to as the base in the discrete choice models, and the model parameters were obtained with PandasBiogeme version 3 (Bierlaire, 2020).

During the estimation of the discrete choice models, scaling procedures were applied for some attributes, which are described in Table 7.

**Table 7. Scaling characteristics of model attributes**

| Attribute | Units | Scaling value |
|---|---|---|
| Detour time | Minutes | 10 |
| Detour cost | UAH | 10 |
| Profit | UAH | 100 |
| Income | UAH | 10,000 |

The estimates of the model based on detour travel costs (DC) and detour travel times (DT) are presented in Table 8. It should be mentioned that the DC for the *car* mode were estimated accounting for detour distances, fuel consumption and fuel price.

**Table 8. Discrete choice modelling results for "cost-time" model**

| Models' coefficients and general statistics | MNL | | | Mixture MNL | | |
|---|---|---|---|---|---|---|
| | *value* | *rob. st. error* | *rob. t-test* | *value* | *rob. st. error* | *rob. t-test* |
| $ASC_{BIKE}$ | −14.09*** | 2.71 | −5.19 | −14.19*** | 2.67 | −5.3 |
| $ASC_{BUS}$ | −0.87** | 0.32 | −2.72 | −0.68* | 0.34 | −1.97 |
| $ASC_{CAR}$ | −9.59*** | 1.74 | −5.52 | −9.77*** | 1.72 | −5.68 |
| $ASC_{ELECTRIC\_GROUND\_PT}$ | −2.62*** | 0.57 | −4.63 | −3.12*** | 0.57 | −5.42 |
| $ASC_{METRO}$ | −2.19*** | 0.72 | −3.05 | −2.97*** | 0.77 | −3.86 |
| $DC_{BUS}$ | −0.63* | 0.32 | −1.95 | −0.67* | 0.34 | −1.97 |
| $DC_{CAR}$ | −4.92*** | 0.91 | −5.42 | −5.08*** | 0.99 | −5.13 |
| $DC_{ELECTRIC\_GROUND\_PT}$ | −2.49*** | 0.34 | −7.37 | −1.87*** | 0.34 | −5.42 |
| $DC_{METRO}$ | −3.16*** | 0.58 | −5.47 | −2.38*** | 0.61 | −3.86 |
| $DT_{BIKE}$ | −21.18*** | 7.23 | −2.91 | −21.00*** | 7.26 | −2.89 |
| $DT_{BUS}$ | −9.82*** | 2.81 | −3.49 | −9.92*** | 2.79 | −3.54 |
| $DT_{CAR}$ | −14.59* | 7.15 | −2.04 | −14.22* | 7.07 | −2.01 |
| $DT_{ELECTRIC\_GROUND\_PT}$ | −7.55*** | 2.26 | −3.34 | −7.50*** | 2.24 | −3.35 |
| $DT_{METRO}$ | −12.35*** | 3.89 | −3.17 | −12.26*** | 3.86 | −3.17 |
| $DT_{WALKING}$ | −4.57*** | 1.23 | −3.71 | −4.57*** | 1.22 | −3.73 |
| $INCOME_{BIKE}$ | 1.28*** | 0.45 | 2.82 | 1.27*** | 0.45 | 2.84 |
| $INCOME_{BUS}$ | 0.28 | 0.26 | 1.06 | 0.26 | 0.26 | 0.96 |
| $INCOME_{CAR}$ | 1.39*** | 0.41 | 3.41 | 1.42*** | 0.41 | 3.44 |
| $INCOME_{ELECTRIC\_GROUND\_PT}$ | 0.743* | 0.34 | 2.12 | 0.75* | 0.34 | 2.19 |
| $INCOME_{METRO}$ | −0.29 | 0.9 | −0.32 | −0.27 | 0.94 | −0.29 |
| $\sigma_{TIME\_BIKE}$ | — | — | — | −0.01 | 0.16 | −0.08 |
| $\sigma_{TIME\_BUS}$ | — | — | — | −0.25 | 0.33 | −0.76 |
| $\sigma_{TIME\_CAR}$ | — | — | — | −0.07 | 0.132 | −0.58 |
| $\sigma_{TIME\_ELECTRIC}$ | — | — | — | 0.01 | 0.069 | 0.22 |
| $\sigma_{TIME\_METRO}$ | — | — | — | 0.09 | 0.062 | 1.55 |
| $\sigma_{TIME\_WALKING}$ | — | — | — | 0.03 | 0.129 | 0.20 |
| *Number of parameters* | 20 | | | 26 | | |
| *Sample size* | 249 | | | 249 | | |
| *Null Log-likelihood* | −431.02 | | | −730.26 | | |
| *Final Log-likelihood* | −182.19 | | | −181.57 | | |
| *Number of draws* | — | | | 100 | | |

Note: robust p-values ***: p < 0.005, **: p < 0.01, *: p < 0.05.

As occasional couriers should cover their cost expenditures for CS deliveries (i.e. additional detours within their routine trips), the profit gained should be considered (Wicaksono et al., 2022). In the context of different modes – which can be divided into travel cost-based and travel cost-free modes (i.e. *bike* and *walking*) – the mode choice behaviour is expected to follow a profit maximization pattern. To account for this aspect, the profit-time MNL and Mixture MNL models have been estimated. The obtained results are summarized in Table 9.

**Table 9. Discrete choice modelling results for the profit-time model**

| Models' coefficients and general statistics | MNL | | | Mixture MNL | | |
|---|---|---|---|---|---|---|
| | *value* | *rob. st. error* | *rob. t-test* | *value* | *rob. st. error* | *rob. t-test* |
| $ASC_{BIKE}$ | −7.49 | 3.92 | −1.91 | −8.34 | 4.88 | −1.71 |
| $ASC_{BUS}$ | 8.21*** | 1.90 | 4.31 | 11.73*** | 2.75 | 4.26 |
| $ASC_{CAR}$ | −7.70* | 3.05 | −2.56 | −8.09* | 3.38 | −2.39 |
| $ASC_{ELECTRIC\_GROUND\_PT}$ | 3.10 | 1.63 | 1.90 | 3.67 | 2.16 | 1.69 |
| $ASC_{METRO}$ | 3.73 | 1.94 | 1.92 | 4.17 | 2.42 | 1.72 |
| $PROFIT_{BIKE}$ | 49.95*** | 13.35 | 3.74 | 61.35*** | 14.33 | 4.28 |
| $PROFIT_{BUS}$ | 56.22*** | 13.26 | 4.24 | 69.31*** | 14.84 | 4.70 |
| $PROFIT_{CAR}$ | 58.70*** | 12.78 | 4.59 | 71.99*** | 14.67 | 4.91 |
| $PROFIT_{ELECTRIC\_GROUND\_PT}$ | 56.62*** | 13.19 | 4.29 | 70.07*** | 14.83 | 4.73 |
| $PROFIT_{METRO}$ | 56.08*** | 13.09 | 4.28 | 69.37*** | 14.68 | 4.73 |
| $PROFIT_{WALKING}$ | 63.98*** | 13.51 | 4.73 | 80.73*** | 15.97 | 5.05 |
| $DT_{BIKE}$ | −31.06*** | 10.25 | −3.03 | −40.36*** | 13.23 | −3.05 |
| $DT_{BUS}$ | −13.67*** | 3.95 | −3.45 | −18.53*** | 5.09 | −3.64 |
| $DT_{CAR}$ | −23.07* | 10.25 | −2.25 | −30.50* | 12.37 | −2.47 |
| $DT_{ELECTRIC\_GROUND\_PT}$ | −10.78*** | 3.19 | −3.37 | −13.83*** | 4.11 | −3.36 |
| $DT_{METRO}$ | −17.85*** | 5.46 | −3.26 | −22.99*** | 7.08 | −3.25 |
| $DT_{WALKING}$ | −6.76*** | 1.78 | −3.79 | −9.01*** | 2.35 | −3.82 |
| $INCOME_{BIKE}$ | 1.55** | 0.57 | 2.72 | 1.80* | 0.7 | 2.54 |
| $INCOME_{BUS}$ | 0.23 | 0.35 | 0.66 | 0.15 | 0.53 | 0.28 |
| $INCOME_{CAR}$ | 1.45*** | 0.50 | 2.88 | 1.57* | 0.64 | 2.45 |
| $INCOME_{ELECTRIC\_GROUND\_PT}$ | 0.72 | 0.40 | 1.76 | 0.85 | 0.57 | 1.50 |
| $INCOME_{METRO}$ | −0.29 | 0.984 | −0.30 | −0.007 | 1.12 | −0.01 |
| $\sigma_{TIME\_BIKE}$ | — | — | — | −0.46* | 0.21 | −2.19 |
| $\sigma_{TIME\_BUS}$ | — | — | — | −0.92*** | 0.31 | −2.94 |
| $\sigma_{TIME\_CAR}$ | — | — | — | 0.15 | 0.14 | 1.08 |
| $\sigma_{TIME\_ELECTRIC\_GROUND\_PT}$ | — | — | — | 0.09 | 0.06 | 1.58 |
| $\sigma_{TIME\_METRO}$ | — | — | — | 0.09 | 0.08 | 1.09 |
| $\sigma_{TIME\_WALKING}$ | — | — | — | 0.41 | 0.23 | 1.83 |
| *Number of parameters* | 22 | | | 28 | | |
| *Sample size* | 249 | | | 249 | | |
| *Null Log-likelihood* | −431.02 | | | −689.42 | | |
| *Final Log-likelihood* | −165.202 | | | −158.63 | | |
| *Number of draws* | — | | | 100 | | |

Note: robust p-values ***: p < 0.005, **: p < 0.01, *: p < 0.05.

Comparing the statistics of the MNL and MMNL models for the cost-time and profit-time cases, it can be seen that accounting for a mixed structure did not significantly improve the final log-likelihood function values. Along with that, $\sigma_{TIME}$ were estimated as not statistically significant for most of modes. Given that, the MNL model estimates are also relevant and can be used for the analysis of mode choice behaviour.

## 5. Discussion

The results indicate that gender heterogeneity is an important aspect within CS delivery service. The analysis revealed that male and female CS couriers behave differently. Women demonstrated a higher preference for the *car* transport mode than men, while men were more willing to use active modes such as *walking* and *bike* for longer distances than women were. Within a range of all modes considered in this study, the findings indicated that female CS couriers are more likely to provide the service with a remuneration 8% lower for the delivery than their male counterparts. This difference is more significant if active modes were used, resulting in an average of 25.45% lower prices for CS services if women would use *bike* or *walking* modes, compared to their male counterparts.

Analysis of acceptable detour times revealed some similarities with the findings obtained by Tapia et al. (2023). For instance, the occasional couriers were ready to implement longer detours – that is, in terms of time – by *car* and *metro* modes than by *bike* (active mode). Given that, similar to the findings by Tapia et al. (2023) it can be stated that the physical effort needed to ride a bike reduced the acceptable detour time for CS deliveries. Considering the gender heterogeneity of potential CS occasional couriers, men and women revealed different behaviour regarding acceptable detour times. The mean acceptable detour time for CS service made by *car* was 41.1 minutes for women compared to 38.4 minutes for men, which means that women appear to be more likely to accept longer detour times than men. Exploring the preferred frequency for providing CS services revealed that slightly over 27% of women respondents were ready to do CS deliveries several times a week, while only 19% of men surveyed considered this frequency acceptable. The difference between female and male couriers was sharper if the frequency patterns were explored with respect to mode choice behaviour. Thus, more than 40% of female respondents indicated their readiness to use a private car for deliveries made several times a week. In turn, only 19% of men couriers consider the *car* mode preferable for deliveries executed several times a week.

In the context of the preferred integration of CS detours into daily routine trips, women revealed their preference to implement CS deliveries within the Work–Home and Work–Groceries trip pairs made by *car*. Men also these trip pairs as preferable for the *car* mode, which allows the authors to assume that CS deliveries in developing economies may generate additional external costs. Particular attention should be paid to the Home–Home (evening time) trip pair, which was considered preferable for CS detour integration, especially among male couriers driving cars. Such behaviour is a clear example of the possible negative effects if CS reaches a wide scale of development and usage. This study found that 38.28% of occasional male couriers would be more likely to implement CS deliveries within their evening activities, which would produce extra trips. As these men couriers revealed their preference to use a *car* for the Home–Home (evening time) trip pair, the CS service would likely place additional pressure on the transport network.

Considering other socio-demographic characteristics, the focus on personal income revealed a strong relationship between couriers' mode choice behaviour and their welfare. For the cost-time model, all $\beta_{INCOME}$ are presented with positive values except for the *metro* mode. Considering the *walking* mode as the base during the discrete choice modelling, a positive value for $\beta_{INCOME}$ indicates an increase in probability of choosing a *car*, *bike*, *electric ground PT* or *bus* instead of *walking*. The highest values of $\beta_{INCOME}$ appeared for *car* mode, and the increase in personal wages would increase *car*-oriented behaviour and contribute to the negative consequences of CS deliveries, as people would likely prefer the *car* mode to provide CS services.

To estimate the possible changes in mode choice due to fluctuations in the attribute values, the marginal probability effect (MPE) (*eq. 6*) was calculated (Boes and Winkelmann, 2006; Schmid and Axhausen, 2019). The considered levels of change for every change in time, cost and profit attributes in the models were: $-10, -5, -1, +1, +5, +10$.

$$MPE_c^a = \frac{1}{N} \sum_{q=1}^{N} \left( P_{qc} \left( mode \mid X^{a^*}; \boldsymbol{\beta} \right) - P_{qc} \left( mode \mid X^a; \boldsymbol{\beta} \right) \right) \qquad (6)$$

where $X^{a*}$ is the model attribute increased by +1%, +5%, +10% or reduced by −1%, −5%, −10%; $X^a$ is the model attribute in base conditions; $\beta$ is the vector of model's parameters; and $N$ is number of observations considered in the study (sample size). MPE was estimated for two models, cost-time and profit-time, to account for the different mode choice behaviours. The estimated results for MPE are presented in Tables 10 and 11, respectively.

**Table 10. Marginal probability effects for cost-time trade-off conditions**

| Estimated attribute | Mode | Percentage change in estimated attribute | | | | | |
|---|---|---|---|---|---|---|---|
| | | −10% | −5% | −1% | +1% | +5% | +10% |
| | | **MPE, %** | | | | | |
| Detour cost, UAH | Bus | 0.43 | 0.21 | 0.04 | −0.04 | −0.21 | −0.41 |
| | Car | 10.12 | 4.49 | 0.74 | −0.66 | −2.67 | −4.24 |
| | Metro | 3.10 | 1.56 | 0.31 | −0.31 | −1.56 | −3.12 |
| | Electric ground PT | 0.87 | 0.42 | 0.08 | −0.08 | −0.40 | −0.78 |
| Detour time, min | Bike | 22.66 | 8.95 | 1.38 | −1.29 | −3.97 | −4.19 |
| | Bus | 39.96 | 13.91 | 1.85 | −1.50 | −5.12 | −7.00 |
| | Car | 21.85 | 12.31 | 1.94 | −1.52 | −5.24 | −10.15 |
| | Metro | 32.42 | 19.86 | 4.78 | −5.15 | −21.89 | −28.49 |
| | Walking | 31.77 | 12.02 | 1.82 | −1.59 | −6.24 | −9.60 |
| | Electric ground PT | 44.95 | 19.65 | 2.18 | −1.60 | −4.71 | −5.79 |

**Table 11. Marginal probability effects for profit-time trade-off conditions**

| Estimated attribute | Mode | Percentage change in estimated attribute | | | | | |
|---|---|---|---|---|---|---|---|
| | | −10% | −5% | −1% | +1% | +5% | +10% |
| | | **MPE, %** | | | | | |
| Profit, UAH | Bike | −3.67 | −2.73 | −0.67 | 0.71 | 4.38 | 12.20 |
| | Bus | −8.84 | −6.88 | −2.08 | 2.56 | 17.40 | 34.74 |
| | Car | −7.89 | −5.36 | −1.44 | 1.63 | 10.03 | 24.11 |
| | Metro | −26.25 | −19.32 | −4.75 | 4.82 | 21.67 | 35.15 |
| | Walking | −12.99 | −10.48 | −2.65 | 2.66 | 12.16 | 20.70 |
| | Electric ground PT | −6.23 | −5.30 | −1.87 | 2.54 | 20.07 | 44.16 |
| Detour time, min | Bike | 27.99 | 11.08 | 1.8 | −1.78 | −4.06 | −4.12 |
| | Bus | 52.04 | 21.59 | 2.71 | −2.07 | −6.45 | −8.34 |
| | Car | 27.61 | 16.91 | 3.46 | −2.65 | −8.10 | −15.71 |
| | Metro | 39.95 | 25.00 | 6.71 | −7.09 | −23.93 | −27.58 |
| | Walking | 26.49 | 9.85 | 1.68 | −1.54 | −6.40 | −9.76 |
| | Electric ground PT | 52.83 | 30.29 | 3.18 | −2.12 | −5.42 | −6.19 |

The MPE for the cost-time model (Table 10) represents the possible changes in mode choice behaviour under common commuting conditions when people may or may not integrate the CS detour into their daily trips. According to the obtained results, the detour time attribute has a notable influence on the mode choice probability, and this effect is more pronounced for PT-based modes such as *electric ground PT*, *metro* and *bus*. Given that, improvements in the transport system (e.g. providing separate lanes for buses) would result in a travel speed increase, which may positively affect *bus* mode choice behaviour. As for detour cost, the MPE estimates for *car* and *metro* modes revealed them to be the most sensitive to cost attributes. Given the trend of increases in in fuel prices, these conditions could lead to the reduction in car usage and a 10% increase in the detour cost is an achievable value in current market trends. In turn, local authorities may increase the fare for metro users (as the metro is a state enterprise in Kharkiv), negatively influencing the preference for *metro* use for CS deliveries. The market prices for electricity may also affect this process.

Accounting for profit-oriented behaviour (Table 11), which is relevant to the scope of CS deliveries, the MPE estimates indicate the relatively high sensitivity to the profit attribute. The higher sensitivity to changes in profit values was determined for travel modes that have a direct travel cost, unlike active modes such *bike*, which is assumed to provide the highest profit. This interesting phenomenon can be explained by the fact that the necessity of accounting for travel costs when the travel is planned produces higher sensitivity to profit-related aspects. For the *bike* mode, this phenomenon is not relevant, as people do not pay for their journey and, in this case, the readiness to provide the CS service is more strongly related to the travel (detour) time aspect. The *walking* mode was revealed to have high sensitivity, especially when the profit values increase. The necessity to carry grocery packages may explain this sensitivity. Given that, the increase in remuneration values paid would reduce the threshold of readiness to provide the CS services while *walking*.

Comparison of the MPE estimates between the cost-time (Table 10) and profit-time (Table 11) models revealed a greater MPE for detour time attributes for profit vs time trade-off conditions. For instance, the usage of the preferred *car* mode by both female and male CS couriers would be reduced under conditions in which the couriers account for values of expected profit. As the *car* mode has the highest detour costs, profit plays a more important role than travel cost within commuting trips (cost-time model). Such assumptions can be proved by analysis of the statistics of the developed models. Given a different number of attributes in the cost-time and profit-time models, the adjusted rho-square criteria were evaluated accounting for degrees of freedom (Koppelman and Bhat, 2006). Having $\overline{\rho}^2_{0(COST-TIME)} = 0.531$ and $\overline{\rho}^2_{0(PROFIT-TIME)} = 0.566$, it can be concluded that there is a better fit of the profit-time model in representing the mode choice behaviour of occasional couriers in the context of e-grocery CS services.

## 6. Concluding remarks

This paper represents the first attempt to explore the mode choice behaviour of occasional CS couriers taking into account their gender heterogeneity. This study considered the multiply mode choice set conditions that CS couriers may face within the transport network of cities. The behavioural data on occasional couriers – such as a preferred mode for CS deliveries, acceptable detour times and expected remunerations for the provided CS service – were collected through an online survey implemented in the city of Kharkiv (1.4 million inhabitants in 2021), Ukraine, in February–March 2021. The sample comprised 249 respondents with a gender distribution of 121 females and 128 males. The survey participants provided identified-preference data on preferred modes for CS deliveries with their specification in terms of spatial and time-related characteristics. To address the issue of unrevealed characteristics for non-chosen alternatives, PTV Visum simulation was employed. The proposed methodology thus provides a full dataset for implementing a mode choice behaviour study based on descriptive and discrete choice analyses.

According to the study results, male and females CS couriers behave differently in the context of the preferred modes for CS deliveries. Female couriers showed a higher threshold for acceptable detours than men did, which reflects their readiness to cover longer distances and provide higher spatial accessibility for CS services. Women couriers also expected, on average, 8% less remuneration for CS deliveries than men did, making them more likely to win bids in the CS digital platform. Despite significant differences in the behaviours of men and women couriers, some common features were identified. For instance, for both genders, *car*-oriented behaviour was found to be dominant. However, in the context of other studied means of transport, women couriers considered active modes such as *walking* and *biking* less attractive for CS deliveries. Male respondents, however, were likely to use the *walking* mode for CS services, as this option was found to have the second greatest share of mal couriers after using their own car. The smallest share of male users was found for the *bus* mode, despite the high accessibility of this mode within the studied transport network. Accounting for the necessity to integrate CS-related trips into daily mobility, women and men couriers considered the Work–Home pair to be preferable. However, in second place women put the travel integration Work–Groceries, while men considered the integration Home–Home (evening) the most relevant after the Work–Home option. This means that male couriers may generate additional trips in the network to perform CS deliveries during the evening, which, in connection with *car*-oriented behaviour, might generate extra external costs and negatively contribute to traffic.

To explore the possible changes in the mode choice preferences related to the spatial, time and cost-related aspects, the average MPE has been estimated. The results indicated that the profit attribute provides the greatest change in the MPE considering the profit–detour time trade-off conditions. This effect is more pronounced for the *metro* and *electric ground PT* modes than for the *car* option, which remained preferable in the current conditions. Given that, it can be supposed that a slight increase in paid remuneration for CS deliveries could make CS deliveries greener and more environmentally friendly, as CS couriers would be more likely to use electric-based means of transport. Within the future planned steps, the authors would like to evaluate the potential externalities of CS services by employing an agent-based modelling methodology. The discrete choice modelling results will be incorporated into the simulation environment to depict the agents' behaviours in a multimodal urban transport system.

### CRediT authorship contribution statement

**Oleksandr Rossolov:** Conceptualization, Methodology, Software, Formal analysis, Investigation, Writing – Original Draft, Editing, Visualization. **Anastasiia Botsman:** Software, Visualization. **Serhii Lyfenko:** Software, Visualization. **Yusak O. Susilo:** Methodology, Formal analysis, Writing – Review & Editing, Funding acquisition.

### Declaration of Competing Interest

The authors declare that they have no known competing financial interests or personal relationships that could have appeared to influence the work reported in this paper.

### Acknowledgement

This work is supported by FFG/BMK Endowed Professor DAVeMoS programme.

**Annex A – Questionnaire**

**CONSENT** to participate in the questionnaire.

**Section 1 SURVEY DESCRIPTION**:

---

*Dear respondents!*

*We are interested in your opinion regarding the online shopping of food products and their direct delivery to consumers. Please fill out the questionnaire below to help us assess the level of interest in this type of shopping and direct delivery of food products under the terms CROWD-SHIPPING.*

*CROWD-SHIPPING is a service for home delivery of food products from a store (supermarket, hypermarket, chain of stores) to the end consumer. Delivery is carried out by occasional couriers using their own vehicles or public transport. The essence of CROWD-SHIPPING is this: food products are delivered to the consumer during a person's main trip as part of his/her daily activities (from home to work, from work to home, etc.). At the same time, a person who expresses readiness to provide such a service does not significantly change his or her daily routine. The delivery is carried out at the same time as trip to the planned destination by an occasional courier. For this service (address delivery), the courier is paid through special mobile applications.*

*The goal of implementing CROWD-SHIPPING is to improve the ecological and socio-economic components of urban space. For example, by involving ordinary people (not commercial firms or delivery companies) to deliver products, the load on city streets for road transport is reduced alongside the reduction in the need for commercial carriers.*

---

Do you currently shop for food online?
- Yes
- No

**Section 2 READINESS TO USE CROWD-SHIPPING SERVICE FOR E-GROCERIES**

---

Imagine that you have decided to buy food products and the following alternatives are available:

**1. ORDER ONLINE IN A STORE OR SUPERMARKET WITH HOME DELIVERY USING CROWD-SHIPPING** (delivery is performed by an individual with whom you will be able to coordinate the delivery process via phone or a mobile application);

**2. ORDER ONLINE IN A STORE OR SUPERMARKET WITH HOME DELIVERY INVOLVING A COMMERCIAL CARRIER** (usually offered by the store);

**3. PURCHASE IN A PHYSICAL STORE.** THE PURCHASE IS MADE IN PERSON.

PURCHASE TERMS: The assortment of food products is the same, regardless of the ordering method: online or during an independent trip (trip) to a physical store (same quality, composition, brands, etc.).

Below you will find nine alternative scenarios. Please make your choice based on your preferences and personal vision of the shopping process on the Internet or in a physical store. The attributes indicated in the proposed options may differ from the reality of your personal life.

Please make your selections based only on the options provided and choose among the three alternative channels that you believe best describe your personal preferences.

Please make ONE CHOICE for each of the NINE SCENARIOS.

---

**Scenario example:**

| Attribute | | Online Commercial Carrier | Online Crowd-shipping Carrier | Physical store |
|---|---|---|---|---|
| Delivery cost | 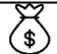 | 70 UAH | 90 UAH | |
| Delivery time | 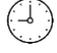 | **6 hours** or less | **3 hours** or less | |
| Ecological contribution | 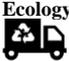 | No reduction of CO2 | Reduction of CO2 | |
| Flexible delivery | 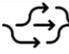 | NO | YES | |
| YOUR CHOICE | | ☐ | ☐ | ☐ |

## Section 3 ATTITUDINAL QUESTIONS

- Rank from 1 to 4 (where 1 is not important and 4 is very important) the importance of the following characteristics of online food shopping with direct delivery on the terms of crowd-shipping.

  o Delivery speed
  o Delivery cost
  o Flexibility of the delivery process
  o Ecological features of delivery

- What is the maximum payment you are willing to pay for home delivery of food products using crowd-shipping?

  o 50 UAH
  o 60 UAH
  o 75 UAH
  o 90 UAH
  o 100 UAH
  o 120 UAH

- How often are you ready to make online food purchases with home delivery using crowd-shipping?
  o Everyday
  o Several times a week
  o Once a week
  o Several times per month
  o Once per month
  o Less than once per month

- Rank from 1 to 5 (where 1 is totally disagree and 5 is totally agree) the importance of the following characteristics:
  o *The cost of online delivery is too high.*
  o *The home delivery saves me time.*
  o *Online shopping is too complicated.*
  o *I like to visit stores and malls.*
  o *The risk of purchasing a low-quality product is the main reason for shopping in store.*
  o *Delivery flexibility like delivery time, place etc. is important for me.*
  o *I am concerned about the security and integrity of the delivery.*
  o *I prefer social contacts when shopping in stores.*
  o *Shopping in stores is very tiring.*
  o *I trust only official representatives.*
  o *The crowd-carrier is more reliable as he is personally responsible for the goods.*

- o *Crowd-shipping allows you to reduce emissions and air pollution.*
- o *I am not concerned about air pollution.*
- o *I travel green for the environment.*
- o *There are more important issues than environmental protection.*

Section 4 **READINESS TO PROVIDE CROWD-SHIPPING SERVICE FOR E-GROCERIES**

*Please answer the following questions to help us evaluate the level of interest in providing home grocery delivery services on the terms of crowd-shipping.*

- The main reason for providing services with crowd-shipping?
  - o Financial motivation
  - o Reduction of traffic due to the exclusion of commercial vehicles from the streets
  - o Ecological improvements to the city's environment

- What minimum remuneration would motivate you to act as a crowd-shipping courier?
  - o 50 UAH
  - o 60 UAH
  - o 75 UAH
  - o 90 UAH
  - o 100 UAH
  - o 120 UAH

- How often would you be willing to provide e-groceries crowd-shipping services?
  - o Everyday
  - o Several times a week
  - o Once a week
  - o Several times per month
  - o Once per month
  - o Less than once per month

- What means of transport do you plan to use for providing crowd-shipping services?
  - o Walking
  - o Private car
  - o Bike
  - o Metro
  - o Tram or trolleybus
  - o Bus

- What is the maximum time you can spend (in addition to your main trip) changing your route and providing related services with crowd-shipping?
  - o 15 min
  - o 30 min
  - o 45 min
  - o 60 min

- When you are able to provide e-groceries crowd-shipping services:
  - o Commuting to work (Home–Work)
  - o Commuting towards home (Work–Home)
  - o Travelling from work for shopping (Work–Groceries)
  - o Within free time outside of work hours (Home–Home)

- What weight are you prepared to carry during a crowd-shipping delivery?
  - o Less than 0.5 kg
  - o From 0.5 to 2 kg
  - o More than 2 kg

- How many crowd-shipping deliveries are you prepared to implement within one day?
    - 1
    - 2
    - 3
    - 4 or more

## Section 5 SOCIO-DEMOGRAPHIC CHARACTERISTICS

***Thank you for your participation in the survey***